\documentclass{article}
\usepackage[T1]{fontenc} % add special characters (e.g., umlaute)
\usepackage[utf8]{inputenc} % set utf-8 as default input encoding
\usepackage{ismir,amsmath,cite,url}
\usepackage{graphicx}
\usepackage{color}

\usepackage{tabularx}

\usepackage{subcaption}

\title{Listen, Read, and Identify:\\Multimodal Singing Language Identification of Music}

\oneauthor
{Keunwoo Choi and Yuxuan Wang}
{ByteDance\\ \tt \{keunwoo.choi, wangyuxuan.11\}\tt@bytedance.com}

\def\authorname{K. Choi}

\begin{document}
\maketitle
\begin{abstract}
We propose a multimodal singing language classification model that uses both audio content and textual metadata. LRID-Net, the proposed model, takes an audio signal and a language probability vector estimated from the metadata and outputs the probabilities of the target languages. Optionally, LRID-Net is facilitated with modality dropouts to handle a missing modality. In the experiment, we trained several LRID-Nets with varying modality dropout configuration and tested them with various combinations of input modalities. The experiment results demonstrate that using multimodal input improves performance. The results also suggest that adopting modality dropout does not degrade the performance of the model when there are full modality inputs while enabling the model to handle missing modality cases to some extent. 

\end{abstract}

\section{Introduction} \label{sec:introduction}
Singing language identification of music (SLID)
is a classification task of labeling tracks by the languages used in lyrics. Language information is essential for music discovery and recommendation systems as 
% understanding of 
lyrics play a crucial role in the music listening experience~\cite{lee2004survey}. In reality, despite its importance, language information is not always available or accurate, even for established, large-scale music streaming services. Such a situation has motivated researchers to develop language identification models using the most fundamental music data -- the audio content itself. 

% For example, it is known that Spotify, an international music streaming service, built an audio-based classifier for 21 languages \cite{roxbergh2019language}. 

There have been various audio-based approaches in SLID. Most of them utilize traditional machine learning classifiers and audio features, similar to early methods for spoken language identification~\cite{sugiyama1991automatic, lamel1994language}. The models in \cite{tsai2004towards} and \cite{tsai2007automatic} consist of vocal/non-vocal segmentation, feature vector quantization, and a simple codebook-based language model.
% 2004:  3 languages. segmentation and feature extraction phoneme classifier:  . 224 tracks. 
% 2007: 346 tracks. same methodology. 
% ``However, singing differs from speech in many ways, including various phonological modifications employed by singers, prosodic shaping to fit the overall melody, and the peculiar wordings used in lyrics. ''\\
% eng and german. pop and rock. two languages, 750+ items per lang, in house, no more information. MFCCs etc. simple source separation  \cite{barry2004sound} did not help. 
% http://ismir2006.ismir.net/PAPERS/ISMIR0699_Paper.pdf
More recently, i-vector \cite{dehak2010front}, a popular feature vector for speech-related tasks, was used in 
\cite{kruspe2014improving, kruspe2014gmm}, combined with a support vector machine classifier. 
% 2014: self-crawled a-capella, 3 languages, 116 to 196 tracks per language.
A vocal source separation technique that is based on a spatial property of stereo music signals was added in a model \cite{schwenninger2006language}, although it did not improve its performance in the experiment. 
More recently, the model in \cite{del2020end} uses a modern vocal separation technique \cite{drossos2018mad} and a one-dimensional deep convolutional neural network.
In these approaches, a vocal source separation module is applied to input signals to extract relevant features with a cleaner signal,  i.e., with a less amount of accompaniments.

% deep 1D conv. 7 languages x 800-1200 tracks in-house dataset, no information specified, f1-score not promising.
% 

There also have been models that are based on non-audio modalities. The model in \cite{chandrasekhar2011automatic} is designed to classify music videos into language categories by taking visual features such as
% principal component analysis and 
histograms of oriented gradients
% that are adapted to video -- 
along with basic audio features such as MFCCs. In the experiment, adding video features improved the accuracy of the model from 44.7\% to 47.8\% in a 25-language classification task. Another model in \cite{roxbergh2019language} uses language estimation of track title and album name and showed a comparable performance to their in-house audio-based classifier. Notably, an internal music representation called track vector, which is estimated using music listening history data, showed the highest feature relevance -- 0.97 -- in their experiment. It re-emphasizes that there is a strong connection between music listening preference and singing language of music. 
% \cite{chandrasekhar2011automatic}: visual + audio features. youtube playlist with language name. 25 languages x 1k tracks per lang. MFCCs etc + HoG etc THEN SVM. MFCC 26\%, all audio 44.7\%, audio and video 47.8\%.  
% http://www.cs.toronto.edu/~dross/ChandrasekharSarginRoss_ICASSP2011.pdf

% Most recently, Roxbergh \cite{roxbergh2019language} using random forest, 20k x 9 languages crawled using Spotify API and language tags coming from the record label. German (de), Spanish (es), Finnish (fi), French (fr), Italian (it), Japanese (ja), Dutch (nl), Portuguese (pt), Swedish (sv).
% selected song name and albume name - 164-dim feature for each country from DetectLanguage (which has limited free api); AND spotify w2v vector (how??), region, etc. f-score 0.952 for macro average. vs audio: hard to conclude due to label noise (no detailed numbers provided), but combining audio and metadata increase the prediction coverage with high confidence!\\
% feature relevance: album name, song name: 0.68 and 0.72. w2v: 0.97 (but limited coverage). 

It is noteworthy that unfortunately, none of the mentioned works is reproducible  \cite{tsai2004towards, tsai2007automatic, schwenninger2006language, chandrasekhar2011automatic, kruspe2014gmm, kruspe2014improving, roxbergh2019language, del2020end} and there has not been any benchmark in SLID. All the previous works rely on private datasets and only little details such as target languages and the number of tracks are known. This is because of a lack of datasets -- there has not been a publicly available music language classification dataset until very recently~\cite{santana2020music4all}. Some music tagging datasets may be considered as alternatives if they include language tags; for example, the million song dataset has language labels~\cite{bertin2011million}. But their tag popularity is merely at 116th (`german'), 135th (`english'), or below, which are often excluded in a prevalent problem formulation such as top-50 classification~\cite{choi2016automatic} to suppress noise during training and evaluation~\cite{choi2018effects}.

In this paper, we introduce LRID-Net -- \textbf{L}isten, \textbf{R}ead, and \textbf{Id}entify-\textbf{Net}work, a model that takes audio and textual metadata (track title, album name, and artist name) to identify singing language. LRID-Net is based on a combination of a deep convolutional neural network, MLPs, and modality dropouts as explained in Section~\ref{sec:proposed}. We provide brief information about Music4All~\cite{santana2020music4all}, the dataset that we use, in Section~\ref{sec:data}. In Section~\ref{sec:experiment}, we present our experiment results including the performance of LRID-Net, a comparison of modalities, and verification of modality dropout. 

\newpage
There are three main contributions of our work.\\
\textbf{Contribution 1}: LRID-Net is the first work in SLID that takes advantage of audio and text inputs,
% (artist name, album name, track title), 
presumably the two most accessible forms of music data.\\
\textbf{Contribution 2}: We present the first reproducible work in SLID by using a public dataset. This enables a rigorous benchmark to be followed, which is necessary for progress in modern machine learning research. \\
\textbf{Contribution 3}: LRID-Net has flexibility in its input data form -- it is designed to work with missing modalities by adopting \textit{modality dropouts}. This flexibility makes LRID-Net a highly pragmatic solution in a real-world scenario.

% Despite its importance, SLID has gained only scarce attention in music information retrieval \cite{tsai2004towards, tsai2007automatic, schwenninger2006language, chandrasekhar2011automatic, kruspe2014gmm, kruspe2014improving, roxbergh2019language, del2020end}. This is because the lack of a public dataset. 

% \footnote{\url{http://www.audiocontentanalysis.org/data-sets/}}

% ------------ Problem -------------- %
\section{Problem Formulation}\label{sec:problem}
% Human listeners can rely on an wide range of information to solve SLID. (Isolated) audio signal may contain lingual cues such as voicing, words, accents, and phonemes. Metadata such..
We define our problem as a \textit{singing language identification using audio content and metadata}. It is reasonable to expect that in many practical use-cases, (some of) three selected types of metadata -- track title, album name, and artist name -- would be accessible. For example, music tracks shared on online streaming services usually include all of them. We exclude other types of data such as visual features and pre-computed track vectors despite their benefits shown in \cite{chandrasekhar2011automatic} and \cite{roxbergh2019language}, respectively, because their availability is limited for a subset of commercial music tracks even for those in industry who have an access to a large-scale proprietary catalog.

To make our model even more practical, we consider a missing data scenario. Some or all of the metadata can be easily missing. For example, indie music tracks shared online (e.g., SoundCloud or Jamendo) usually do not include any album name, or they can exist but should be considered missing since titles can be blank or consists of numbers and/or special characters only, not providing any linguistic information. The audio could be also missing or not helpful at all, for example, a segment that is input to the model may not contain any vocal part.

From a machine learning point of view, our SLID problem is a single-label multi-class classification with a multi-modal, potentially partially missing input. In fact, some lyrics are multilingual. But we assume that those cases are negligible, following the dataset we use (see Section~\ref{sec:data} for more details).\footnote{English words such as ``Yeah'', ``Hey", and ``Baby" are often used as musical expressions or interjections in non-English songs and we do not assume their existence makes a lyric multilingual.}
% The same assumption was made during the creation of Music4All dataset (see Section~\ref{sec:data} for more details).

% This problem formulation inspires us to answer the following research questions - \textbf{RQ 1} -- \textbf{4}.\\
% \textbf{1}. How well does the proposed multimodal model work?\\
% \textbf{2}. Can modality dropout cope with a missing modality?\\
% \textbf{2b}. Is there any disadvantage by considering missing modality use-cases?\\
% \textbf{4}. Would a vocal source separation improve the model?

% ------------ Dataset -------------- %
\section{Dataset}\label{sec:data}
We use Music4All dataset \cite{santana2020music4all} which includes 30-second audio clips (44,100 kHz and stereo), lyrics, and 16 other metadata such as title, album name, artist name, and Spotify identifier of 109,269 tracks. The dataset also includes language labels covering 46 languages. They are estimated from the lyrics using Langdetect,\footnote{\url{https://pypi.org/project/langdetect/}} a Python implementation of Language-Detection~\cite{nakatani2010langdetect}.\footnote{We noticed there are errors in the ground truth. But Langdetect is known to perform at 99\% accuracy with documents, which is significant higher than the performance of the proposed audio-based models.}

The distribution of language labels in Music4ALL is heavily skewed from 84,103 items (English) to 1 item (Hindi, Slovak, Bulgarian, and Hebrew).
% (see Table~\ref{table:langdetect_detail} for details) 
We consider 11~labels -- top-10 popular languages and an ``others" category that includes all the other languages. In detail, there are 84,103 English tracks followed by Portuguese (7,020), Spanish (3,225), Korean (1,145), Others (1,059), French (994), Japanese (615), German (577), Polish (446), Italian (437), and Slovakian (231). There are also 9,417 instrumental tracks, but we exclude them because otherwise, the model would need to learn to perform language identification as well as instrumental track classification, which would distract our analysis.

There does not exist an official training-validation split of tracks of M4A dataset. 
% We use a 80:20 split proposed in  
We use an 80:20 stratified split with allocating every artist in only one of the split sets.\footnote{\url{https://github.com/keunwoochoi/music4all_contrib}} This prevents artist-dependent information from being a confounding factor and leaking the information between training and validation sessions.

\section{LRID-Net}\label{sec:proposed}

\begin{figure}[t!]
  \centering
 \includegraphics[width=1.0\columnwidth]{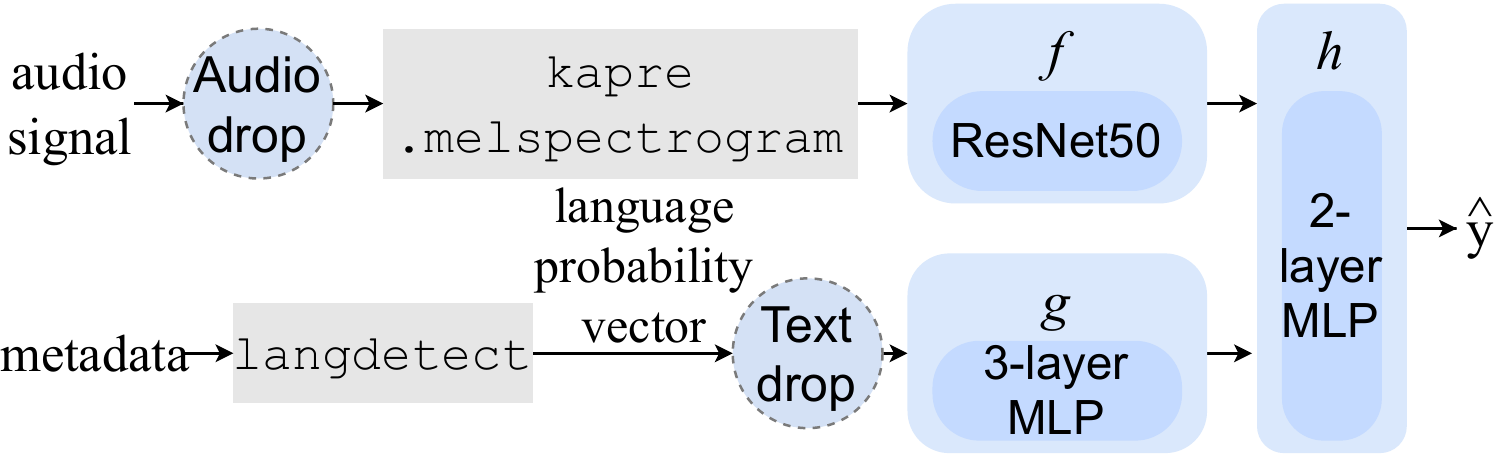}
 \caption{Block diagram of LRID-Net. Gray rectangular boxes indicate non-trainable modules and dotted circles are modality dropouts (Section~\ref{subsec:explain_modality_dr}), which are optionally applied in some of the experiments (Section~\ref{subsec:modality_dr}).}
\label{fig:diagram}
\end{figure}

\subsection{Input Audio Preprocessing}
The 44.1~kHz~sampled 30-second stereo audio signals are downmixed and resampled to 22,050~Hz and converted into 128-bin log-magnitude mel-spectrograms (128~$\times$~2580) with 1024-point FFT and 256 hop size using Kapre~\cite{choi2017kapre}. We call this spectrogram $x_{\text{audio}}$. 

\subsection{Input Text Preprocessing}
The metadata strings are joined in an order of artist name, album name, and track title and then input to Langdetect to estimate a language probability vector. Originally, \texttt{langdetect.detect\_langs()} outputs a probability distribution of 55 supported languages. However, there are some cases where the function fails to estimate probability, e.g., if the text is blank or a numeric value only. We add another dimension to indicate those exceptions. This 56-dimensional vector is called $x_{lang}$.
% todo; we should discuss later that by doing so, there is effectively already some (static) dropout is happening. 

\subsection{Overall model structure}
LRID-Net consists of two input branches that are concatenated at a late stage of the network.

\textbf{The audio branch, $f$}, is a ResNet-50 with 64 base channels and outputs a 2048-channel feature map that is a size of (4 $\times$ 81)~\cite{he2016deep}. We choose ResNet-50 for its simplicity yet strong performance as shown in~\cite{won2020evaluation} for music tagging. Then, a global average pooling is applied to output a one-dimensional vector length of 2048. This procedure is represented as $s_{\text{audio}}=f(x_{\text{audio}})$.

\textbf{The text branch, $g$},
is a 3-layer MLP where each layer consists of a 128-unit fully-connected layer, a batch normalization layer, and a ReLU activation \cite{ioffe2015batch}, i.e., $s_{\text{text}}=g(x_{\text{lang}})$. 

\textbf{In the output branch, $h$}, 
the two outputs of the input branches are concatenated, i.e., $s_{\text{cat}}=[s_{\text{audio}}; s_{\text{text}}]$ and input to an MLP. This MLP consists of a 256-unit fully-connected hidden layer, a batch normalization layer, a ReLU activation, and an 11-unit fully-connected layer with a Softmax activation to output the language probability, i.e.,  $\hat{y}=h(s_{\text{cat}})$.

\subsection{Modality Dropout}\label{subsec:explain_modality_dr}
We use \textit{modality dropout} which was originally introduced in \cite{neverova2015moddrop} as \textit{ModDrop} and is illustrated in the block diagram in Figure~\ref{fig:diagram}. Similar to the original dropout~\cite{neverova2015moddrop}, during training time, a modality dropout module replaces its input with zeros with a probability of $r$, the dropout rate. There are two main differences between the original dropout and modality dropout. 1) The original dropout is applied to a part (e.g., a single node or a channel) of an input but a modality dropout module drops the whole input of a modality, i.e., an audio signal or a language probability vector. By doing so, it effectively simulates a missing modality input and lets the model learn to perform the task without the dropped input. 2) There is no $1/(1-r)$ scaling in a modality dropout when the input is not dropped. During test time, a system with LRID-Net inputs a zero vector to the model if a modality is missing.

% ------------ Experiment -------------- %
\section{Experiment and Discussion}\label{sec:experiment}
We performed a series of experiments to demonstrate the performance and properties of LRID-Net. We use Music4All dataset~\cite{santana2020music4all} and process audio and text data using Kapre~\cite{choi2017kapre} and Langdetect~\cite{nakatani2010langdetect} as detailed in Section~\ref{sec:data} and Section~\ref{sec:proposed}, respectively. During training, we use Adam optimizer~\cite{kingma2014adam} and early stopping with a patience of 20 epochs. We do not adopt any balancing during batching and loss computation.

On the metric, we use F1-score, precision, and recall.
As described in Section~\ref{sec:data}, there is a high imbalance of the number of data points of each language in the dataset. In this case, from a user perspective, a macro average can be used to represent the class-balanced performance to avoid the bias towards popular languages, too. However, \textit{because} they are biased in the \textit{same way} in the training and validation sets, weighted (or micro) averaging can be considered to be more suitable than macro averaging on representing how successfully the model was trained to minimize the empirical loss. Acknowledging this issue, we use both of the averaging methods and focus on the performance with a suitable one depending on the context.

We present the performance of each language sorted by its occurrence count, as known as Support, to help understanding of any related trend.

\subsection{Langdetect baseline} \label{subsec:exp_langdetect}

\begin{table}[t]
\begin{tabular}{lrrr}
\small
\textbf{Text input} & \multicolumn{1}{l}{\textbf{Precision}} & \multicolumn{1}{l}{\textbf{Recall}} & \multicolumn{1}{l}{\textbf{F1-Score}} \\ \hline
Artist Name         & .323                                   & .221                                & .149                                  \\ \hline
Album Name          & .399                                   & .378                                & .284                                  \\ \hline
Track Title               & .450                                   & .444                                & .317                                  \\ \hline
Joining All    & \textbf{.510}                                   & \textbf{.569}                                & \textbf{.429}                                 
\end{tabular}

\caption{The macro averaged performance of Langdetect prediction with various text inputs. `Joining All' represents the performance of langdetect baseline.}
\label{table:langdetect_overall}
\end{table}

\begin{table}[t]
\begin{tabular}{lrrr}
\small
\textbf{Text input} & \multicolumn{1}{l}{\textbf{Precision}} & \multicolumn{1}{l}{\textbf{Recall}} & \multicolumn{1}{l}{\textbf{F1-Score}} \\ \hline
Artist Name         & .600                                   & .368                                & .456                                  \\ \hline
Album Name          & .750                                   & .576                                & .652                                  \\ \hline
Track Title               & .766                                   & .573                                & .656                                  \\ \hline
Joining All    & \textbf{.922}                                   & \textbf{.819}                                & \textbf{.857}                                 
\end{tabular}

\caption{The weighted averaged performance of Langdetect prediction with various text inputs. `Joining All' represents the performance of Langdetect baseline.}
\label{table:langdetect_overall_wa}
\end{table}

\begin{table}[t]
\resizebox{\columnwidth}{!}{%

\begin{tabular}{lrrrr}
\textbf{Language} & \multicolumn{1}{l}{\textbf{Precision}} & \multicolumn{1}{l}{\textbf{Recall}} & \multicolumn{1}{l}{\textbf{F1-Score}} & \multicolumn{1}{c}{\textbf{Support}} \\ \hline
English           & .969                               & .859                                & .911                                  & 84,103                               \\ \hline
Portuguese        & .897                                   & .695                                & .783                                  & 7,020                                \\ \hline
Spanish           & .676                                   & .620                                & .647                                  & 3,225                                \\ \hline
Korean            & .444                                   & .003                                & .007                                  & 1,145                                \\ \hline
Others            & .092                                   & .700                                & .162                                  & 1,059                                \\ \hline
French            & .367                                   & .585                                & .452                                  & 994                                  \\ \hline
Japanese          & .895                                   & .153                                & .261                                  & 615                                  \\ \hline
German            & .093                                   & .773                                & .166                                  & 577                                  \\ \hline
Polish            & .695                                   & .572                                & .627                                  & 446                                  \\ \hline
Italian           & .229                                   & .748                                & .350                                  & 437                                  \\ \hline
Slovakian         & .252                                   & .554                                & .347                                  & 231                                 
\end{tabular}
}
\caption{The performance of Langdetect prediction when all the metadata are concatenated (`Joining all'). Support column indicates the number of items in the training set, of which distribution is preserved in the validation set.}
\label{table:langdetect_detail}
\end{table}

\begin{figure}[t]
  \hspace{-0.9cm}
 \includegraphics[width=1.1\columnwidth]{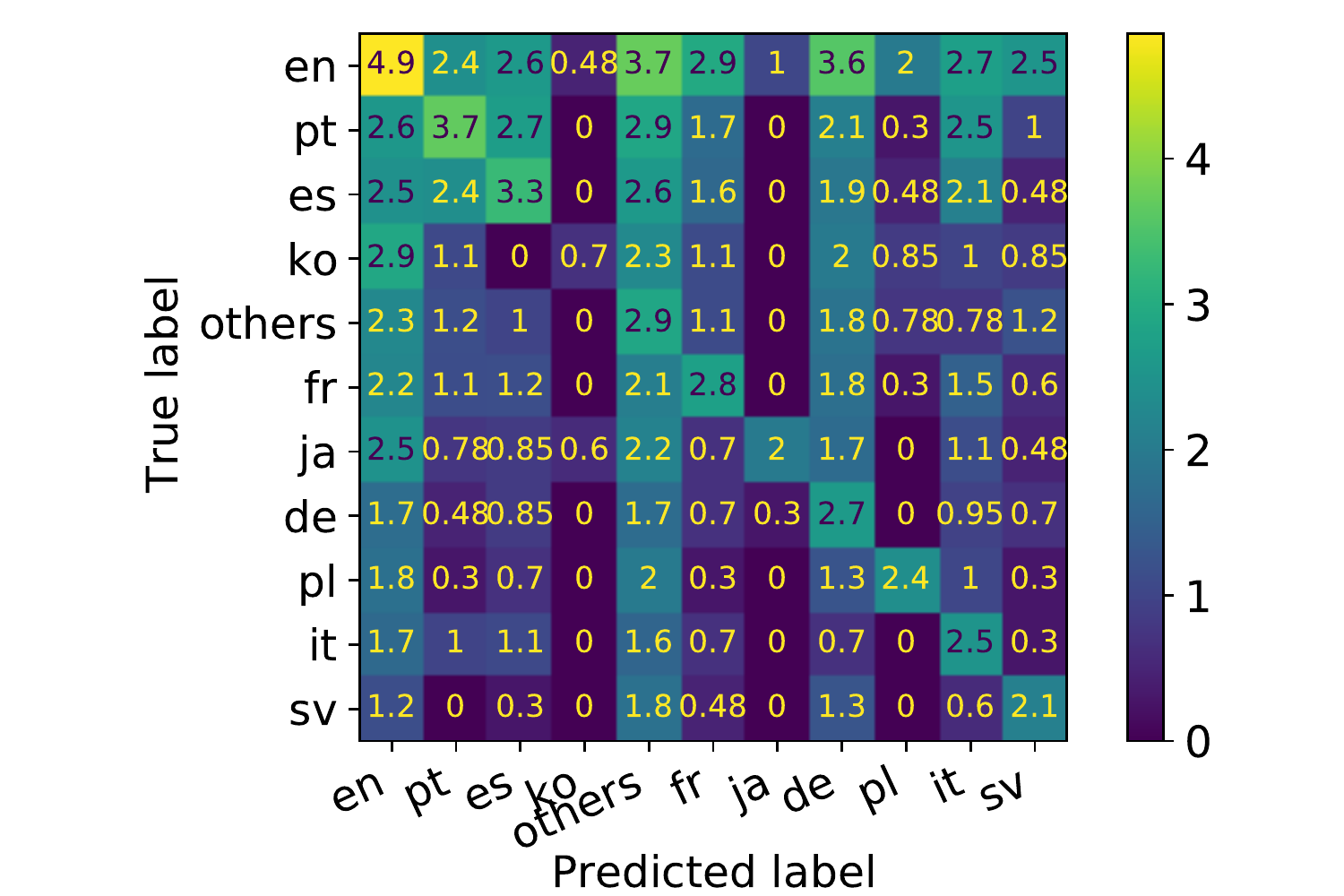}
 \caption{The confusion matrix of Langdetect baseline model (`Joining all'). Item counts are converted by $log_{10}(x+1)$.}
\label{fig:langdetect_confusion}
\end{figure}

We present the result of a simple solution that is to directly use the top prediction of text-based language identification using Langdetect. This baseline approach is called \textit{Langdetect baseline}. We also present a detailed analysis of the performance and behavior of Langdetect baseline model to deepen our understanding of the problem and the following experiment results.

Table \ref{table:langdetect_overall} and \ref{table:langdetect_overall_wa} summarize the performance of Langdetect baseline based on various text inputs, showing that using all the metadata (`Joining All') performs the best by achieving an F1-score of 0.429 (macro average) or 0.857 (weighted average). 
Details of the performance with `Joining All' are presented in Table \ref{table:langdetect_detail} and its confusion matrix is illustrated in Figure~\ref{fig:langdetect_confusion}. Note that Langdetect baseline model is not trained with our dataset, hence support (number of true items) does not affect the performance. 

The F1-scores seem under-performing since Langdetect is reported to show 99\% accuracy. We conjecture two reasons for this result.
First, music metadata is significantly shorter than typical documents and news article, with which Langdetect was originally trained and tested, respectively. 
Second, there is a prevailing usage of English for artist name, album name, and track title even if the lyrics are not written in English, especially for those songs that are internationally consumed. 

In detail, the precision and recall shows interesting patterns that are partly related to the second reason; There are languages where precision is significantly higher than recall (Group 1: Korean, Japanese) while precision is significantly lower than recall for some other languages (Group 2: French, German, Italian, Slovakian). We note that the languages in Group 1 are only two non-European languages and suggest two explanations for their patterns. First, on high precision scores, since Korean and Japanese use completely different letter systems compared to other languages, it is very easy for Langdetect to recognize them if the metadata is written in Korean or Japanese. Second, the low recall may come from the common usage of English as mentioned earlier. 
% perhaps because of such a difference, it is extremely common to use English in metadata among successful K-pop and J-pop artists and 
In such cases, metadata-based Langdetect would never be able to correctly predict that the lyrics are written in Korean or Japanese. This is revealed in a deeper dataset analysis. Among the 1,145 Korean songs in the training set, there are 244 unique artists, out of which 241 artist names are English.
% \Minz{MW: Currently, it looks like an assumption. Maybe it would be better to verify it by checking their characters of the missing cases.} 
% \Minz{MW: Also Japanese songs have way higher scores than Korean songs although there are more Korean songs in the dataset. I suspect this is because many Korean songs with English metadata are written in alphabets while Japanese write them in Katakana.} --> skipping explaining this (no space, perhaps it's a too much detail)
Conversely, the model almost never misclassifies non-Korean or non-Japanese songs to Korean or Japanese, respectively. 
% Beside English (which shows a high precision/recall) and those in Group 1, the model mutually confuses songs with the languages in Group 2.
Unlike Group~1, we did not find any convincing explanation for the patterns of Group~2.
% This is also clearly illustrated in the confusion matrix in . The mode rarely classify any item to be Korean or Japanese, either correctly or incorrectly. We do not suspect this is an issue about the software, \texttt{langdetect}, given the completely differentiating character sets of Korean and Japanese languages from other languages.

\subsection{Single modality baselines} \label{subsec:fx_multimodal}
% The comparison in Section \ref{subsec:exp_mainmodel} explains many aspects of the proposed Main model. However, there are some limitations because i) the text part of Main model is different from Langdetect baseline and ii) it does not provide how much a model can perform with audio signals only. 
As additional baseline models, we show experiment results of single modality models. They are AO (audio-only) model and TO (text-only) model.

\begin{figure}[t]
\small
 \begin{subfigure}{1.0\columnwidth}
  \centering
  \includegraphics[width=1.0\columnwidth]{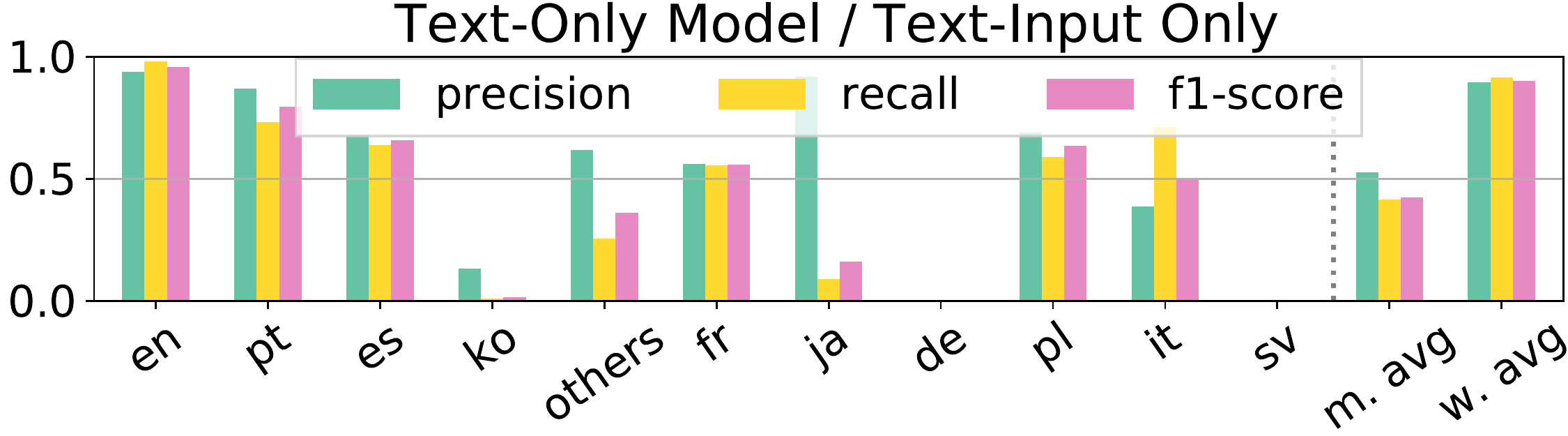}
 \end{subfigure}
 \begin{subfigure}{1.0\columnwidth}
  \centering
 \includegraphics[width=1.0\columnwidth]{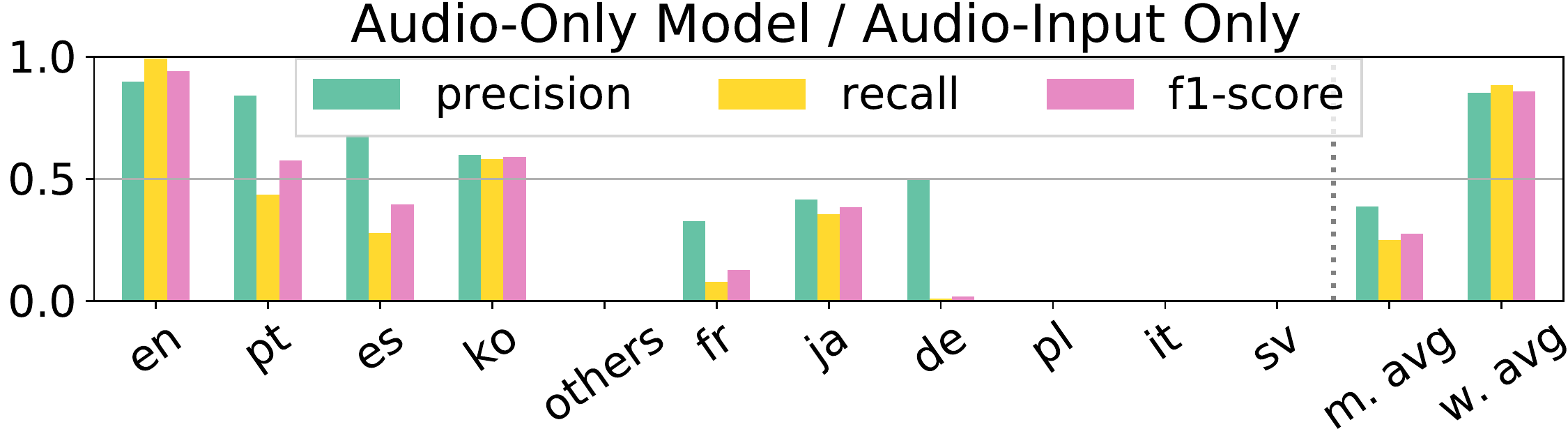}
 \end{subfigure}
 \caption{Top - The performance of an text-only model (TO~model). Its precision, recall, and F1-score are .526~/~.415~/~.422 (macro averaging) and .896~/~.914~/~.900 (weighted averaging). Bottom - The performance of an audio-only model (AO~model). The precision, recall, and F1-score are respectively .387~/~.248~/~.275 (macro averaging) and .852~/~.884~/~.857 (weighted averaging).}
\label{fig:exp20_result}
\end{figure}

First of all, we compare the TO model against Langdetect baseline. TO model achieves a comparable macro averaged F1-score (0.422, a degradation of 0.007) and a higher weighted averaged F1-score (0.900, an improvement of 0.043) as in Figure~\ref{fig:exp20_result}. Again, this result seems heavily affected by the class imbalance of the training set.

Second, as illustrated in Figure \ref{fig:exp20_result}, a lack of a modality leads to negative effects, often critically to some languages. AO model completely fails at identifying Others, Polish, Italian, and Slovakian while TO model fails at Korean, Japanese, German, and Slovakian.

Third, as summarized in Figure~\ref{fig:exp20_result}, in every metric and averaging strategy, TO~model outperformed AO~model, showing the importance of using metadata. However, this does not mean audio is less useful than textual data. 
Acknowledging the limit of the information in the metadata discussed with Langdetect baseline in Section~\ref{subsec:exp_langdetect}, the result may indicate the opposite that currently, the information from text input is almost saturated and more improvement should be based on a better audio understanding.
% assuming that the groundtruth of the dataset (language estimation from lyrics matched by audio IDs) is highly reliable, 

% this comparison shows that the current audio-based approach has a room to improve.

\subsection{LRID-Net: Main Model}\label{subsec:exp_mainmodel}

\begin{figure}[t]
  \centering
 \includegraphics[width=1.0\columnwidth]{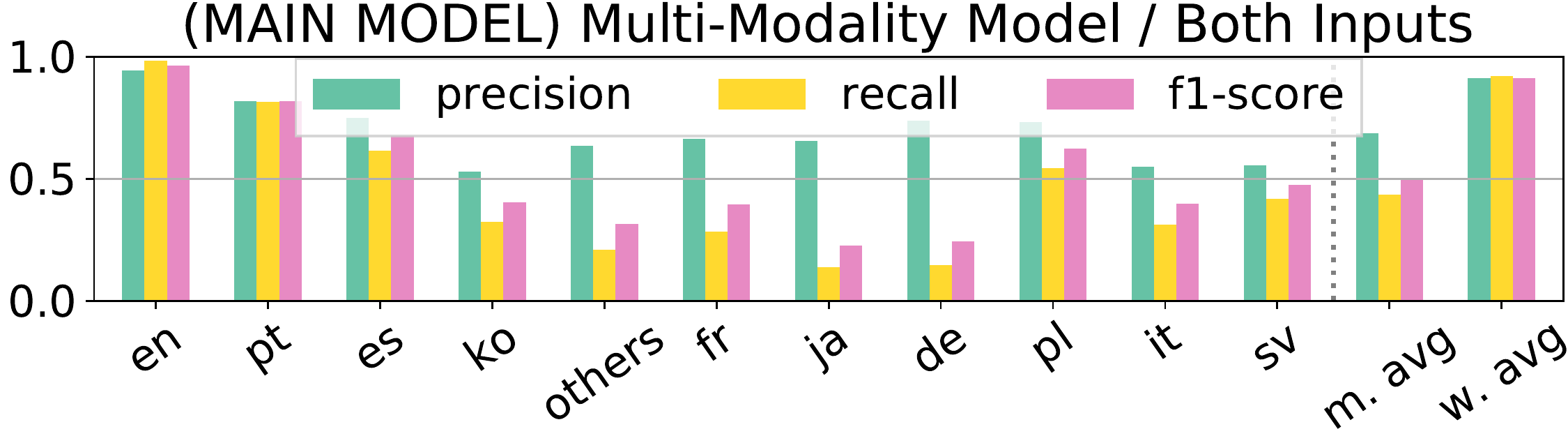}
 \caption{The performance of Main model with both modality inputs. The precision, recall, and F1-score are respectively .688~/~.435~/~.504 (macro averaging) and .911~/~.922~/~.911 (weighted averaging).}
\label{fig:exp10_result_bar}
\end{figure}

\begin{figure}[t]
%   \centering
\hspace{-1cm}
 \includegraphics[width=1.05\columnwidth]{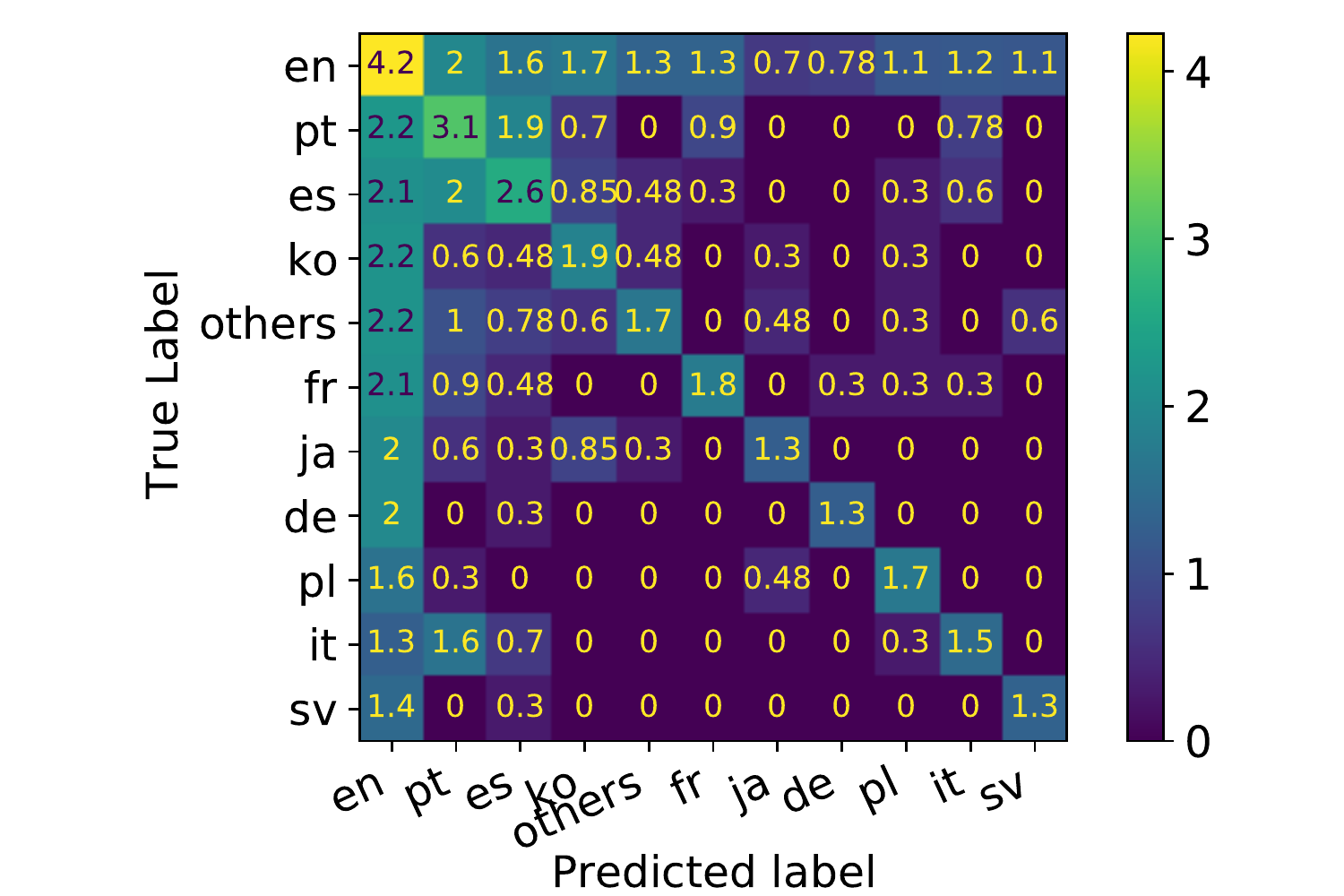}
 \caption{The confusion matrix of Main model. Item counts are converted by $log_{10}(x+1)$.}
\label{fig:exp10_result_confu}
\end{figure}

In this section, we introduce the experiment result of our multimodal SLID model, LRID-Net. This LRID-Net net was trained without any modality dropouts and we call this model `Main model'.

Main model shows an improvement over all the aforementioned baseline models. Compared to Langdetect baseline, it achieves a higher F1-score (+0.048) with a higher precision (+0.134) and a lower recall (-0.118) on weighted average. It also achieves a higher F1-score (+0.075) and a higher precision (+0.158), but a lower recall (-0.134) on macro average. Main model also outperforms AO~model and TO~model, the single modality models, in every metric and averaging strategy, emphasizing the benefit of using multimodal information. 
% Overall, it shows that the training is successful and an LRID-Net can perform the task, utilizing multimodal inputs. % We can expect a more balanced result if the distribution is balanced in the training set, too.\\ 

Among languages, Main model shows low recall rates for Korean, French, Japanese, German, and Italian as shown in Figure~\ref{fig:exp10_result_bar}. This pattern is similar to that in the results of Langdetect baseline as discussed in Section \ref{subsec:exp_langdetect}. We conjecture that a similar type of confusion may have happened in Main model, especially if the 56-dimensional text-based language probability input computed with Langdetect plays an important role (which seems true given that TO~model outperforms AO~model in Section~\ref{subsec:fx_multimodal}). The class imbalance of the training set seems to penalize recall rates of those languages because a classification of unconfident items is likely to be biased towards the mode of the distribution of training items, i.e., English. This is shown in Figure~\ref{fig:exp10_result_confu}.

The overall improvement of F1-score did not benefit all the languages equally. The performance of Main model is rather polarized than Langdetect baseline model. The F1-scores of 3/4 popular languages (English, Spanish, and Korean) and Others category are improved in Main model compared to Langdetect baseline. In the meantime, out of the six less popular languages, only two languages (Italian and Slovakian) show an improvement. This might mean a correlation between the performance and the number of training data.
% This can be partially explained by the class imbalance in the training set, but not completely. 
However, there is a clear exception. Main model achieves an F1-score of 0.347 for Slovakian, which both AO~model and TO~model completely failed.

For some languages, Main model achieved a lower performance than a single modality model, leaving room for further improvement. For example, Main model is outperformed by AO~model for Korean and TO~model for French and Italian. 
% Third, the effect of different modality usage is not homogeneous among languages. For example, in terms of F1-score, Portuguese, Spanish, and Slovakian benefit from multimodality while Main model is outperformed by AO~model for Korean and TO~model for French and Italian. Both TO~model or AO~model completely fails at classifying Slovakian, the least popular language in our dataset.

\subsection{Modality Dropout} \label{subsec:modality_dr}
 
In this section, we discuss the benefits and effects of applying modality dropout to Main model. We test three modality dropout strategies: applying modality dropout to audio input only (ADr-Main model), text input only (TDr-Main model), and both of the inputs (ATDr-Main model). The dropout rate is fixed to 0.2 for every model.\footnote{0.2 was chosen assuming a small portion (for example, 20\%) of tracks would have missing modality in the real use-cases.} In ATDr-Main model, the two dropouts work independently. This means that during training, stochastically, 4\% of training items would have zero values for both of the inputs where backpropagation is still applied to update the model. This leads to strengthen the model to predict the distribution of the training data and may not be ideal, but we did not observe any critical issue in practice.

Since there is no language-specific pattern, we present the averaged metrics only in this section.

\subsubsection{Case 1: Complete Modality -- Do modality dropouts have any negative affects?}
\begin{figure}[t]
\small
  \centering
 \includegraphics[width=1.0\columnwidth]{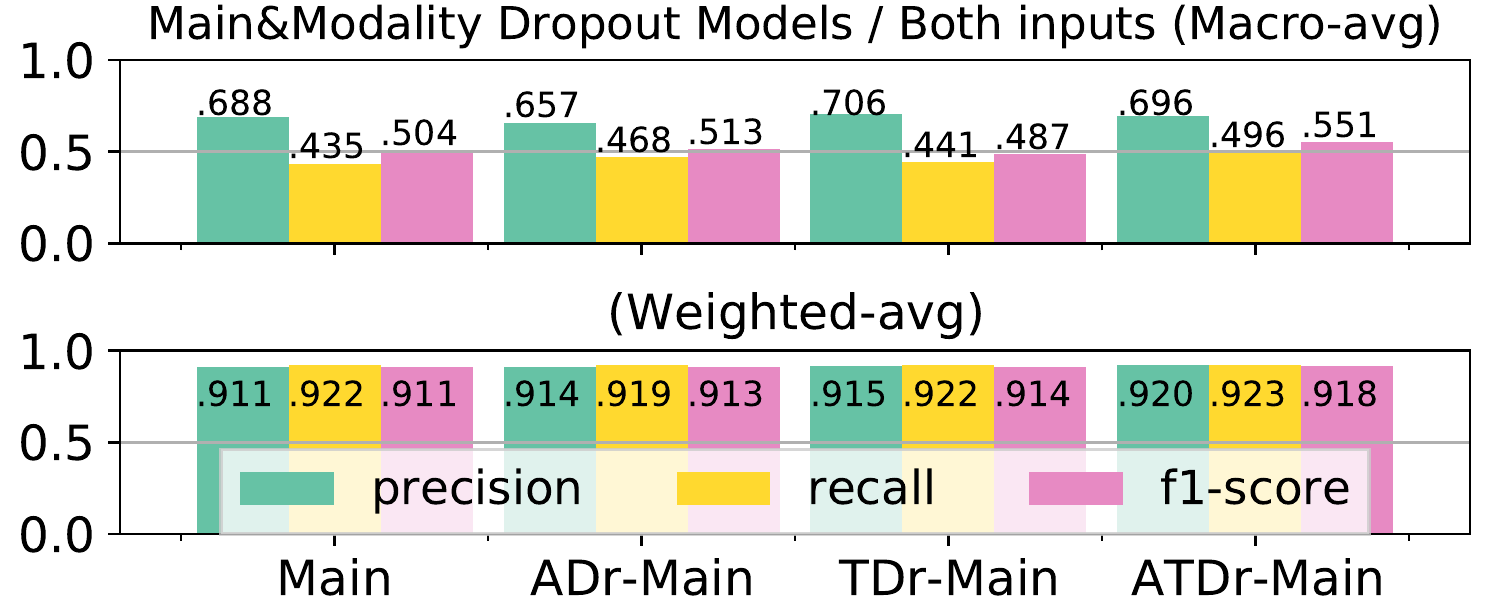}
 \caption{The comparison of metrics of Main model and models with various modality dropout strategies when there is no missing modalities in the data. ADr-Main, TDr-Main, and ATDr-Main indicate models with the same architecture but with modality dropouts applied on audio input, text input, and both of the inputs, respectively.}
\label{fig:exp30_result}
\end{figure}

It is unusual to use modality dropout in music information retrieval. We first investigate to ensure that adopting it does not harm the normal use-cases, i.e., when there is no missing modality.

As presented in Figure \ref{fig:exp30_result}, all the models with modality dropouts - ADr-Main model, TDr-Main model, and ATDr-Main model achieve comparable or even outperforming performances over Main model although they might not be statistically significant. The effects of Modality dropout are better reflected on weighted-average scores than macro-average ones. That is because weighted-averaged scores are more linearly related to the empirical loss that our models are trained to minimize. As shown, modality dropouts only improved those scores. To summarize, adopting modality dropouts does not harm the normal use-cases.

\subsubsection{Case 2: Missing Audio Input} \label{subsec:missing_mode_audio}

\begin{figure}[t]
\small
  \centering
 \includegraphics[width=1.0\columnwidth]{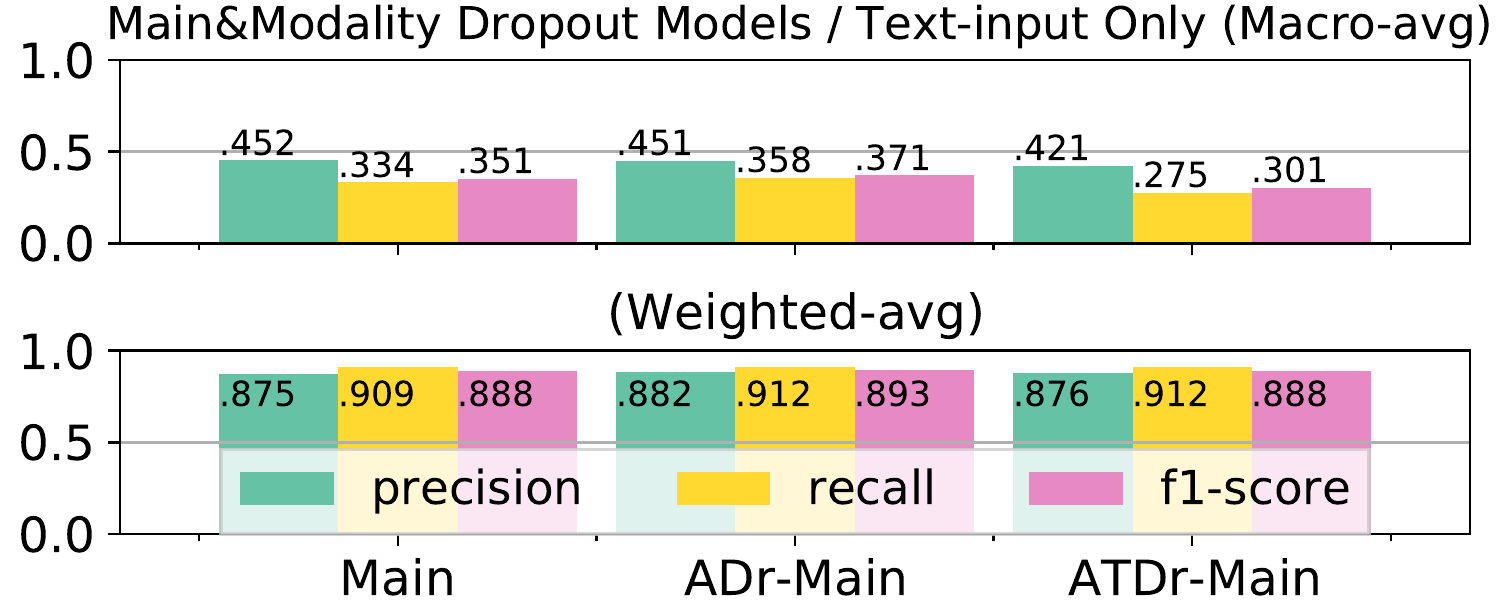}
 \caption{The comparison of metrics of Main model and models with various modality dropout strategies (see \ref{fig:exp30_result} for details) when there is missing audio modality. Note that as in Figure~\ref{fig:exp20_result} (top), TO~model, a dedicated text-only model achieved precision / recall / F1-score of .526 / .415 / .422 (macro averaging) and .896 / .914 / .900 (weighted averaging), respectively.
 }
\label{fig:exp30_audioless}
\end{figure}

Figure~\ref{fig:exp30_audioless} presents the performances of Main model, ADr-Main model, and ATDr-Main model when audio inputs are missing, i.e., there is only text input. 

First, surprisingly, Main model - one that is trained without any modality dropout - performs comparably with ADr-Main model and ATDr-Main model, which are trained explicitly to be able to deal with missing audio input. This is an unexpected behavior since there is no reason the model should learn to ignore a zero audio input (silence) and make a correct prediction solely based on the text information -- but, seems like it does. We find it difficult to explain it and leave it as a future work. 
% \Minz{MW: Didn't you have any dropout in your h(x-cat)? If you have dropout there, it can behave pretty similar to the modality dropout I guess.}
% --> probably not imo?

Second, it is worth comparing these results with that of TO~model (Figure~\ref{fig:exp20_result}, top), a model that is designed and trained to work with text input only. In all the metrics and averaging strategies, TO~model outperforms all the three models. For F1-score, even with the best model (ADr-Main), there is a difference of 0.051 (macro averaging) or 0.007 (weighted averaging). This indicates that despite versatility, a model with an audio modality dropout may not completely replace a dedicated text-only model, especially if missing audio inputs are highly likely.

\subsubsection{Case 3: Missing Text Input}

\begin{figure}[t]
\small
  \centering
 \includegraphics[width=1.0\columnwidth]{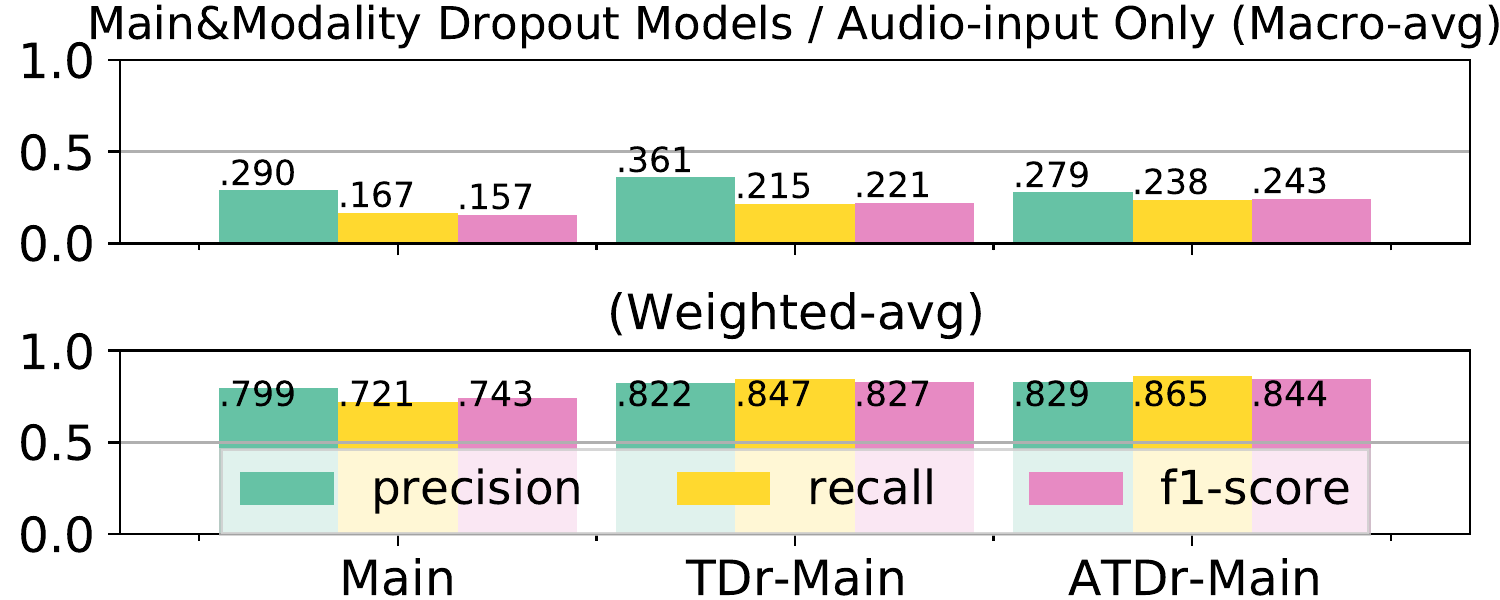}
 \caption{The comparison of metrics of Main model and models with various modality dropout strategies (see \ref{fig:exp30_result} for details) when there is missing text modality. 
 Note that as in Figure~\ref{fig:exp20_result} (bottom), AO~model, a dedicated audio-only model achieved precision / recall / F1-score of 387 / .248 / .275 (macro averaging) and .852 / .884 / .857 (weighted averaging), respectively.}
\label{fig:exp30_textless}
\end{figure}

Figure~\ref{fig:exp30_textless} presents the performance of Main model, TDr-Main model, and ATDr-Main model when text input is missing, i.e., there is only audio input.

First, there are noticeable improvements by applying modality dropouts - TDr-Main~model and ATDr-Main~model outperformed Main~model for the most of the metrics. ATDr-Main~model achieved +0.086 and +0.101 higher F1-scores than Main model does. This result supports a potential real-world use-case of serving a single SLID model where metadata may or may not be available. 

Second, when compared to AO~model (Figure~\ref{fig:exp20_result}, top), all the three models in Figure~\ref{fig:exp30_textless} are outperformed. This means, similar to the conclusion of Section~\ref{subsec:missing_mode_audio}, a model with a text modality dropout may not serve as a perfect alternative and whether to apply text modality dropout (as opposed to train two different models, AO~model and Main~model) would be a practical choice: the decision would be based on the ratio of missing-text inputs and the costs of training and maintaining one vs. two models.

\section{Conclusion}
In this paper, we presented LRID-Net, a deep learning model for singing language identification (SLID) that takes advantage of multimodal data. LRID-Net takes an audio input as well as a text input that combines track title, album name, and artist name. We also propose modality dropout in MIR task, which is designed to let a single model be used with varying input availability. In the experiment, we showed that i) multimodal input improves the performance, ii) a language probability vector of metadata is an effective representation for SLID, iii) modality dropouts do not harm the performance when both of the input modalities exist, and iv) modality dropouts make a model robust with missing input to some extent.

Our research has several limitations. There are some behaviors that we could not provide a satisfying explanation about. Although being useful to some extent, the modality dropout did not completely fulfill the need of building multiple models to cope with missing modalities.
% There would be a better method and parameters to simulate missing modality cases than the modality dropout used in the experiment. 
Due to the already complicated experiment configuration, we did not opt for balancing the languages, which would be a necessary step to build a more practical language classifier.

% In general, SLID is a relatively less explored area and 
There are many research questions to be answered in SLID. A data-driven SLID model might learn some non-linguistic features that are correlated to language labels, and identifying those mechanisms would lead to building more robust SLID models. One approach to demystifying their behavior is to use source separation techniques and observe how much a model is replying on vocal parts vs accompaniments. 
Source separation techniques also could lead to a better performing model because separated signals would provide a disentangled input representation that may be useful for the task. 
Finally, another highly related and interesting task is lyric transcription, which can be combined with SLID, where a mutual benefit is anticipated.

% don't forget to comment this out in the ismir submission!

\section{Acknowledgement}
We thank Jeong Choi, Minz Won, Jordan Smith, Janne Spijkervet, Gianluca Micchi, and Zhihao Ouyang for their helpful reviews and discussion on this paper.

\bibliography{reference}

\end{document}